\documentclass[11pt]{report}                                               
\usepackage{palatino}
\usepackage[english]{babel}
\usepackage{amsmath, amsthm,amssymb}
\usepackage{float}
\usepackage{rotating}
\usepackage{amsmath}
\usepackage{fancyhdr}
\usepackage[normalsize]{caption}

\usepackage{footmisc}
\usepackage{epsfig}
\usepackage{graphicx}
\usepackage{color,hyperref}
    \catcode`\_=11\relax
    \newcommand\email[1]{\_email #1\q_nil}
    \def\_email#1@#2\q_nil{%
      \href{mailto:#1@#2}{{\emailfont #1\emailampersat #2}}
    }
    \newcommand\emailfont{\sffamily}
    \newcommand\emailampersat{{\small@}}
    \catcode`\_=8\relax    

\newcommand{\bs}{\bigskip}

\newcommand{\bc}{\begin{center}}
\newcommand{\ec}{\end{center}}

\usepackage{relsize}

\begin{document}

\begin{flushright}
SLAC-PUB-15291 \\
BABAR-PROC-12/066 \\
November, 2012 \\
\end{flushright}

\bs

\bc 
{\huge \bf{Measurement of Collins asymmetries in inclusive production
of pion pairs in $e^+e^-$ collisions at BaBar}}

\bs

\underline{Isabella Garzia}~ (on behalf of the BaBar Collaboration)

\bs

INFN-Sezione di Ferrara,\\
SLAC National Accelerator Laboratory, Menlo Park, CA\\
E-mail: \email{garzia@slac.stanford.edu}
 
 \ec

\bs

\begin{center}
\large \bf Abstract
\end{center}
We present a preliminary measurement of the Collins asymmetries
in the inclusive process $e^+e^-\rightarrow q\bar{q}\rightarrow \pi^{\pm} \pi^{\pm} X$ at 
center-of-mass energy near 10.6 GeV.
We use a data sample of 468~fb$^{-1}$ collected by the BaBar experiment,
and we consider pairs of charged pions produced in opposite jets
in hadronic events.
We confirm a non-zero Collins effect as observed by previous 
experiments, and we study the Collins asymmetry
as a function of pion fractional energies and transverse momenta,
and as a function of the polar angle of the analysis axis.

\vspace{3cm}
\vbox{\raggedright
                           \it 
                         Proceedings for the  36th International Conference on High Energy Physics,\\
		July 4-11, 2012\\
		Melbourne, Australia}

\newpage

\section*{Introduction}

Transverse spin effects in fragmentation processes were first discussed by 
Collins~\cite{collins}, who introduced the chiral-odd polarized
fragmentation function $H_1^\perp$, also called the Collins function, 
which describes the distribution of the final state hadrons 
produced by the fragmentation of a transversely polarized quark
relative to the momentum direction of their quark.

The first experimental evidence of a non-zero Collins function was obtained
in semi-inclusive deep inelastic scattering (SIDIS)~\cite{sidis}, 
where the convolution of the Collins function 
and the parton transversity distribution function $h_1$
was measured\footnote{The transversity parton distribution
function $h_1$ is a poorly known chiral-odd function.
The first extraction of $h_1$~\cite{global}
was made possible only after the independent measurement 
of the Collins effect by the Belle Collaboration~\cite{belle}.
}.

\begin{figure}[!htb]
\centering
 \includegraphics[width=0.38\textwidth] {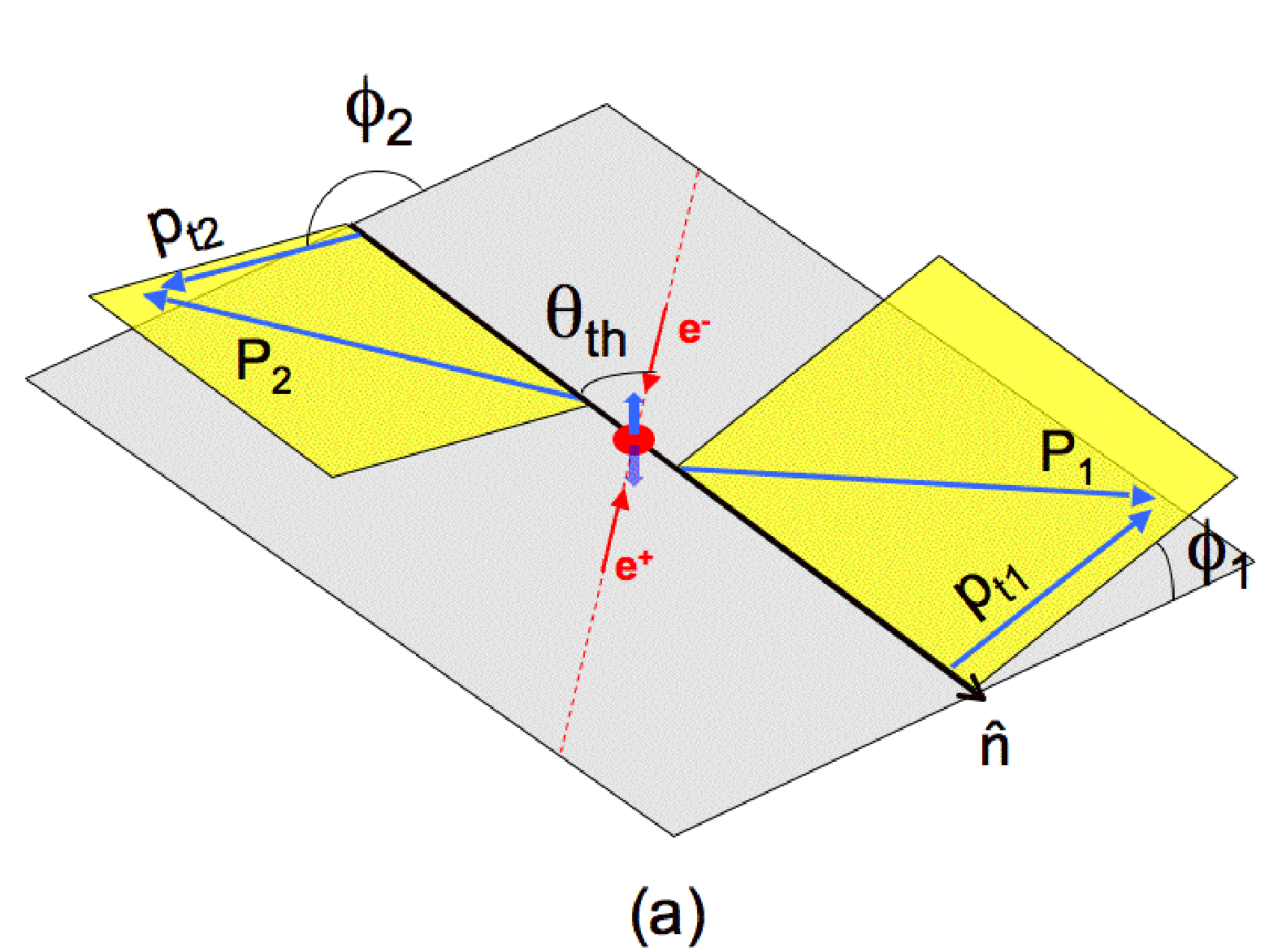}
  \includegraphics[width=0.38\textwidth] {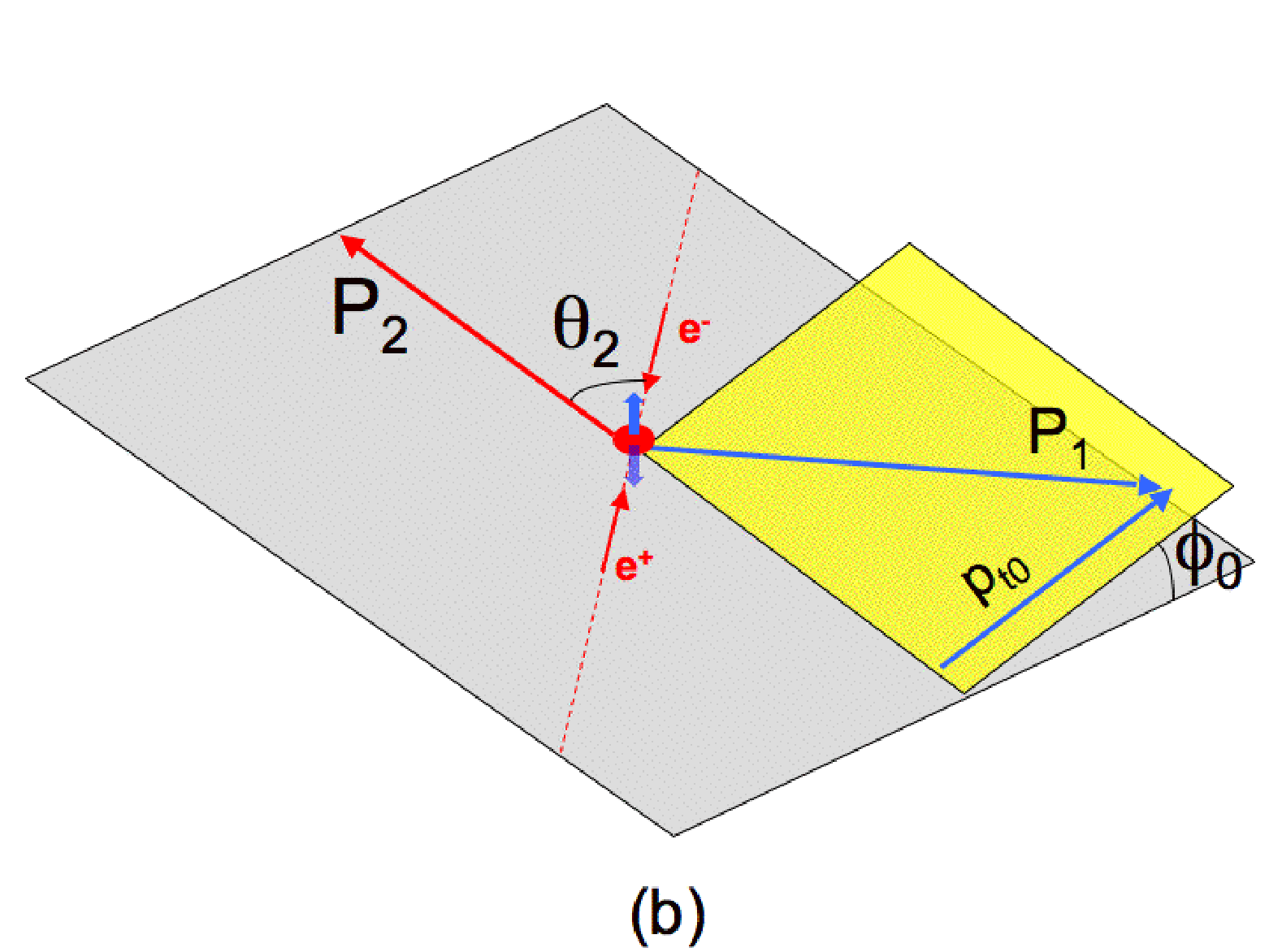}
\caption{ (a) Thrust reference frame or RF12: $\theta=\theta_{th}$ is the angle between the
 $e^+e^-$ collision axis and the thrust axis ($\hat{n}$)~\cite{thrust}, 
$\phi_{1,2}$ are the azimuthal angles between the scattering plane 
and the  momentum transverse to the thrust axis, $\mathbf{p}_{t1,t2}$.
Note that the thrust axis provides a good approximation to the $q\bar{q}$ axis, 
so that $\mathbf{p}_{ti}\simeq \mathbf{p}_{\perp i}$   in Eq.~(1.1).
(b) Second hadron frame or RF0: $\theta_2$ is the angle 
between the collision axis and the second hadron momentum; 
$\phi_0$ is the azimuthal angle between the plane defined by the 
beam axis and the second hadron momentum $P_2$, 
and the first hadron's transverse momentum $\mathbf{p}_{t0}$.
All  tracks are boosted to the $e^+e^-$ center of mass frame.
}
\label{fig:sdr}
\end{figure}

Direct information on the Collins function can be obtained 
from $e^+e^-$ annihilation experiments via the study of the 
semi-inclusive processes $e^+e^-\rightarrow q\bar{q}\rightarrow \pi\pi X$,
where two charged pions coming from the fragmenting 
$q\bar{q}$ pairs ($q=u,\, d,\, s$) 
are detected.
In $e^+e^-$ annihilation,
the measurement of the Collins asymmetry can be performed 
in two reference frames~\cite{boer}, as described in Fig.~\ref{fig:sdr}.
We refer to them as the thrust reference frame or RF12 (Fig.~\ref{fig:sdr}(a)), and
the second hadron reference frame or RF0 (Fig.~\ref{fig:sdr}(b)).

The cross section in the $e^+e^-$ center of mass frame is proportional to
\begin{equation} 
\sigma \sim 1+ \sin^2\theta \cos\phi \frac{H_1^\perp(z_1,\mathbf{p}_{\perp1})\overline{H}_1^\perp(z_2,\mathbf{p}_{\perp2})}{D_1^\perp(z_1,\mathbf{p}_{\perp1})\overline{D}_1^\perp(z_2,\mathbf{p}_{\perp2})},  \label{eq:cross}
\end{equation}
where $D_1$ is the well-known unpolarized fragmentation function, 
the bar denotes the $\bar{q}$ fragmentation, $z$ is the pion fractional energy,
$\mathbf{p}_{\perp}$ is the transverse momentum of the pions with respect to the $q\bar{q}$
direction, $\theta$ is the polar angle of the analysis axis,
and $\phi$ is a proper combination of the pion azimuthal angles
($\phi_1+\phi_2$ in the RF12 frame, or $2\phi_0$ in the RF0 frame).

The $\cos\phi$ modulation in Eq.~(\ref{eq:cross}), which
multiplies the product (convolution) of two $H_1^\perp$ in the RF12 (RF0) frame,
produces an azimuthal asymmetry around the
quark momentum, called the Collins effect or Collins asymmetry.
The first independent measurement of the Collins effect
 was performed by the Belle Collaboration~\cite{belle},
while the first simultaneous extraction of $H_1^\perp$ and $h_1$ was
obtained by the authors of Ref.~\cite{global}.

We report the measurement of the Collins asymmetry in $e^+e^-$ annihilation
based on a data sample of $468$ fb$^{-1}$ collected by the BaBar experiment~\cite{babar}
at  center-of-mass energy   $\sqrt{s}\sim10.6$ GeV.
We also perform a new measurement of the asymmetry as a function
of the transverse momentum $p_t$ of pions
with respect to the analysis axis.

\section*{Analysis strategy}

Assuming the thrust axis~\cite{thrust} to define the $q\bar{q}$ direction and selecting pions
in opposite hemispheres with respect to the thrust axis,
we measure the azimuthal angles $\phi_1$, $\phi_2$, and $\phi_0$.
In order to select the two-jet topology, an event thrust 
value larger than $0.8$ is required.
Only pions coming from the primary vertex with a fractional energy
$z=2E_\pi/\sqrt{s}$ in the range between $0.15$ to $0.9$ are selected.
The Collins asymmetries can be accessed by measuring
the $\cos\phi$ modulation of the normalized distributions of the selected pions
in the two reference frames.
The asymmetries resulting from these distributions are significantly affected by
detector acceptance effects, making their measurement unreliable.
We therefore create suitable ratios of asymmetries in order to eliminate
detector effects and first order radiative effects~\cite{boer}.
We construct different ratios by selecting different combinations
of charged pions: pions with same charge (L), opposite charge (U),
and the sum of the two samples (C).
In this way we also access information on the favored and 
disfavored fragmentation functions~\cite{global}, where a favored
process describes the fragmentation of a quark of flavor $q$
into a hadron containing a valence quark of the same flavor.
These ratios are fitted with a function linear in $\cos\phi$,
\begin{equation} \label{eq:dr}
\frac{N^{U}(\phi_i)/<N^U>}{N^{L(C)}(\phi_i)/<N^{L(C)}>}=B_i+A_i\cdot\cos\phi_i,
\end{equation}
where $i=12$ or $i=0$ identifies the reference frame (RF12 or RF0),
 $\phi_{i}=\phi_1+\phi_2$ or $\phi_i=2\phi_0$, $N(\phi_i)$
  is the di-pion yield, $<N>$ is the average bin content,
and $A_i$ is the parameter sensitive to the Collins effect.
Thanks to the large amount of data, corresponding to 
about $10^9$ events, we are able to choose a $6\times6$ ($z_1,z_2$) 
matrix of intervals,
with boundaries $z_i=0.15,\, 0.2,\, 0.3,\, 0.4,\, 0.5,\, 0.7,$ and $0.9$,
and  the following $p_t$ intervals: $p_t<0.25$ GeV/c,
$0.25<p_t<0.5$ GeV/c, $0.5<p_t<0.75$ GeV/c, and $p_t>0.75$ GeV/c.

\section*{Study of systematic effects}

A crucial point for the measurement of the Collins asymmetry
is the identification of all systematic effects that can influence
the azimuthal distributions of the pion pairs.
We test the double ratio method on a Monte Carlo (MC) sample.
In addition, we study the influence of particle identification
and the uncertainties due to the fit procedure,
the possible dependence of detector response on pion charge,
the presence of residual polarization of the beams, and other minor effects.
When the effects are sizable we correct the measured asymmetries
for them and assign appropriate systematic errors.
All systematic uncertainties and/or corrections are evaluated 
for each interval of fractional energy $z$ and transverse momentum $p_t$.

\subsection*{Dilution of the asymmetries due to the thrust axis}

In this analysis we approximate the $q\bar{q}$ axis by the thrust axis.
The distribution of the opening angle between the two axes 
peaks at about 100 mrad, and this deviation 
 leads a dilution of the measured asymmetry.
This effect can be evaluated using a MC sample. 
However, spin effects, and so the Collins fragmentation function,
are not implemented in our JETSET generator~\cite{jetset}.
Therefore, we re-weight the angular distributions
of the generated tracks (that is before the detector simulation is applied),
in order to reproduce the expected asymmetry\footnote{
Note that for the generated sample
the analysis axis is the true $q\bar{q}$ axis in the RF12 frame,
and is the generated momentum of the second pion in the RF0 frame.}.
The reconstructed azimuthal distributions of the re-weighted 
sample are then fitted, and the resulting asymmetries are compared 
to the applied weights.
We find that the asymmetries are significantly underestimated in the RF12 frame,
and we evaluate the correction factors for each
range of $z$ and $p_t$ according to the measured value.

\subsection*{Contribution of background events to the asymmetries}

The presence of background processes can introduce 
azimuthal modulations not related to the Collins effect.
The background sources giving a significant contribution after 
the selection procedure are:
$e^+e^- \rightarrow \tau^+\tau^-$, $e^+e^-\rightarrow c\bar{c}$,
and $e^+e^-\rightarrow B\overline{B}$, and we refer to them as 
$\tau$, charm, and bottom backgrounds, respectively.
\begin{figure}[!htb]
\centering
 \includegraphics[width=0.45\textwidth] {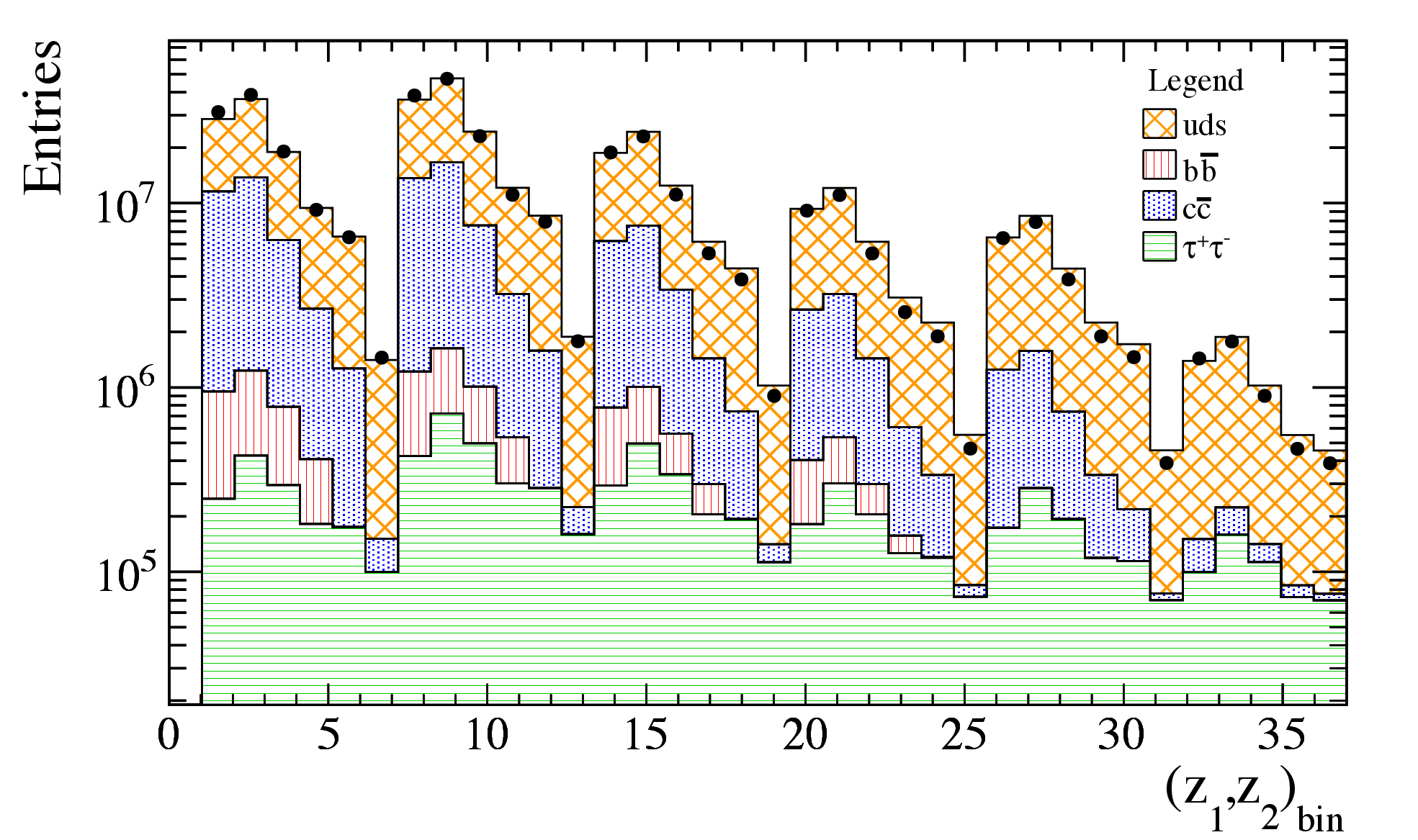}
\caption{Fraction $F_i$ of pion pairs due to $\tau^+\tau^-$,
$B\overline{B}$, $c\bar{c}$, and $uds$ components 
estimated using MC simulation as a function of
$6\times6$ ($z_1,z_2$) matrix intervals.
The black dots represent the fraction $F$ in the data sample.
The same distribution is obtained for $f_i$ in the $D^*$-enhanced data sample.}
\label{fig:molt}
\end{figure}
The asymmetry measured ($A^{meas}$) by fitting the double ratio 
of Eq.~(\ref{eq:dr}) includes also the azimuthal dependence 
of every physics process contributing to the final sample,
and can be written as:
\begin{equation}\label{eq:bg}
A^{meas}=(1-\sum_i F_i) \cdot A^{uds}+\sum_i F_i \cdot A^i,
\end{equation}
where $F_i$ and $A^i$ are respectively the asymmetry and the 
fraction of pion pairs due to the $i^{th}$ background component,
with $i=\tau$, charm or bottom.
The fraction $F_{bottom}$ is very low (less than $2\%$), 
while $F_\tau$ is relevant only for very energetic tracks.
In addition, we measured $A_\tau$ in a $\tau$-enhanced data sample,
and found that the asymmetry is consistent with zero.
For these reasons, in Eq.~(\ref{eq:bg}), we set $A_\tau$ and $A_{bottom}$
equal to zero.
The charm contribution is the dominant background,
with $F_{charm}\sim 30 \%$ on average; both
fragmentation processes and weak decays can introduce 
an azimuthal modulation.
To estimate this contribution
we select a charm-enhanced data sample requiring at least
one $D^*$ candidate from the decay
$D^{*\pm}\rightarrow D^0\pi^\pm$, with the $D^0$ candidate
reconstructed in one of the following four decay channels:
$D^0\rightarrow K^-\pi^+$, $D^0\rightarrow K^-\pi^+\pi^-\pi^+$,
$D^0\rightarrow K^0_S\pi^+\pi^-$, and $D^0\rightarrow K^-\pi^+\pi^0$.
Given $A^{meas}$ in the full data sample and $A^{D^*}$ in the $D^*$-enhanced
sample, we can write the following equations:
\begin{eqnarray}
A^{meas}&=&(1-F_{charm}-F_{\tau}-F_{bottom})\cdot A^{uds}+F_{charm}\cdot A^{charm}\label{eq:meas}\\
A^{D^*}&=&f_{charm}\cdot A^{charm}+(1-f_{charm}-f_{bottom})\cdot A^{uds}. \label{eq:dstar}
\end{eqnarray}
Inverting these equations, we extract the real
contribution from light quarks ($A^{uds}$) to the Collins asymmetry.

A significant source of systematic error in this procedure 
can arise from the fractions $F_i$ ($f_i$) shown in Fig.~\ref{fig:molt}, which are estimated 
using MC samples.
We conservatively  assign as errors
the observed data-MC differences 
 to the most significant contributions
($F_{charm}, F_{\tau}$, and $f_{charm}$),
and we propagate them together with the statistical errors
through Eqs.~(\ref{eq:meas}),~(\ref{eq:dstar}).

\section*{Preliminary results and conclusion}

\begin{figure}[!htb]
\centering
 \includegraphics[width=0.55\textwidth] {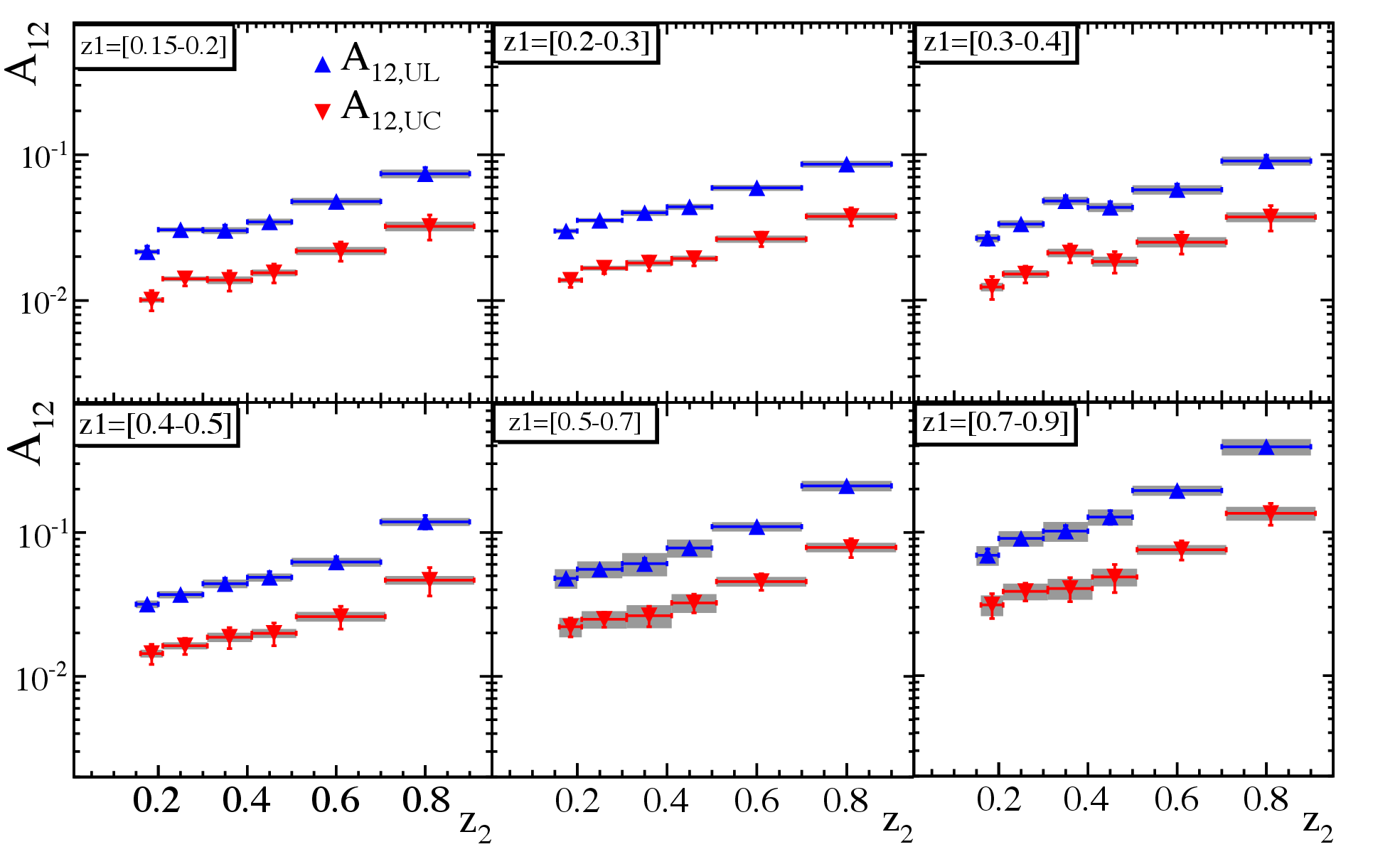}
 {\boldmath
      \put(-180,163){{(a) BaBar Preliminary}} 
    }
  \includegraphics[width=0.4\textwidth] {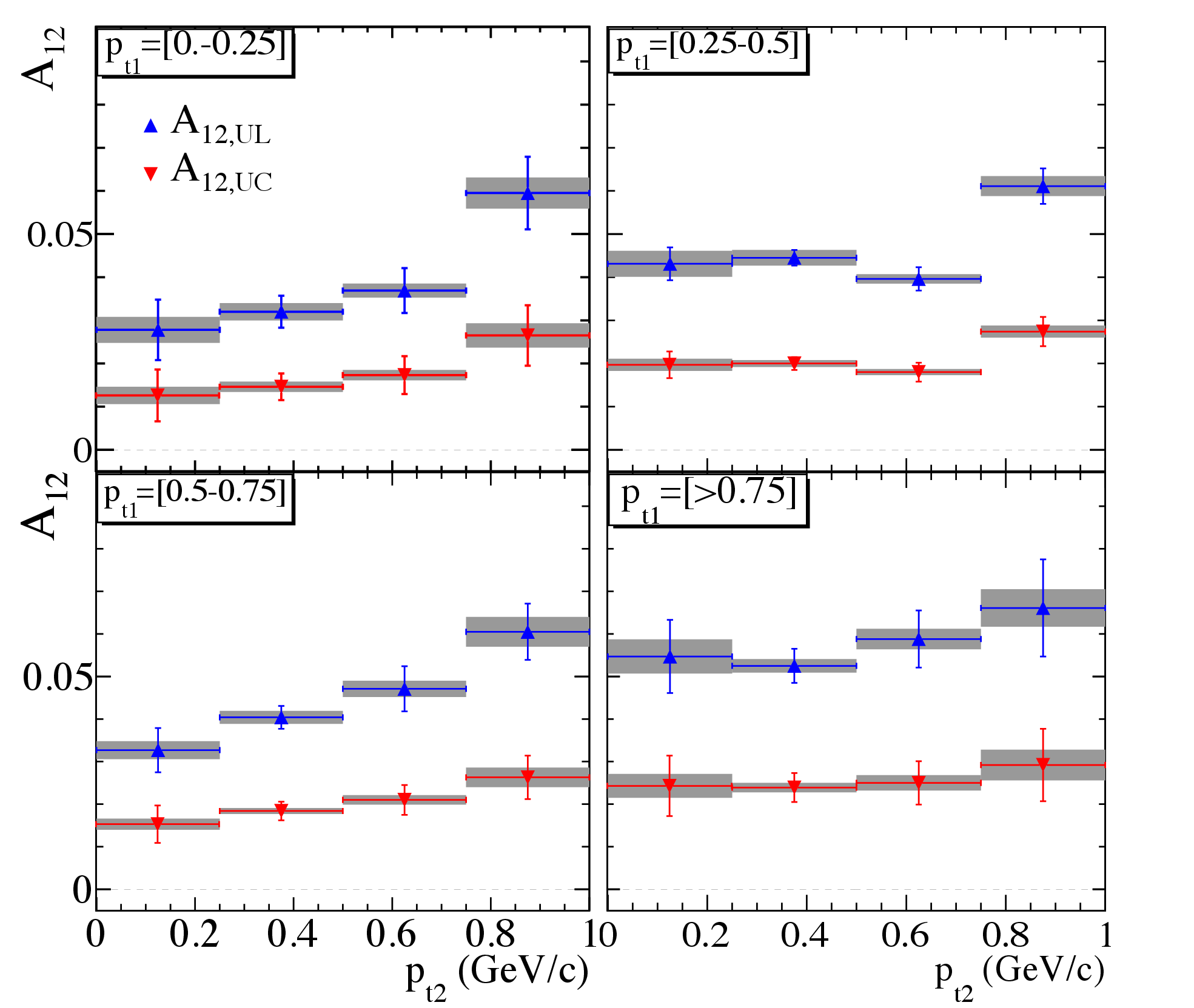}
   {\boldmath
      \put(-145,163){{(b) BaBar Preliminary}} 
    }
\caption{RF12 frame: (a) Collins asymmetries for light quark as a function of 
$(z_1,z_2)$ , and (b) as a function of $(p_{t1},p_{t2})$ intervals.
Blue triangles refer to the ratio U over L (UL), while red triangles
to the UC ratio.
Statistical and systematic errors are shown as error bars and bands
around the points, respectively. 
}
\label{fig:results}
\end{figure}

\begin{figure}[!htb]
\centering
 \includegraphics[width=0.48\textwidth] {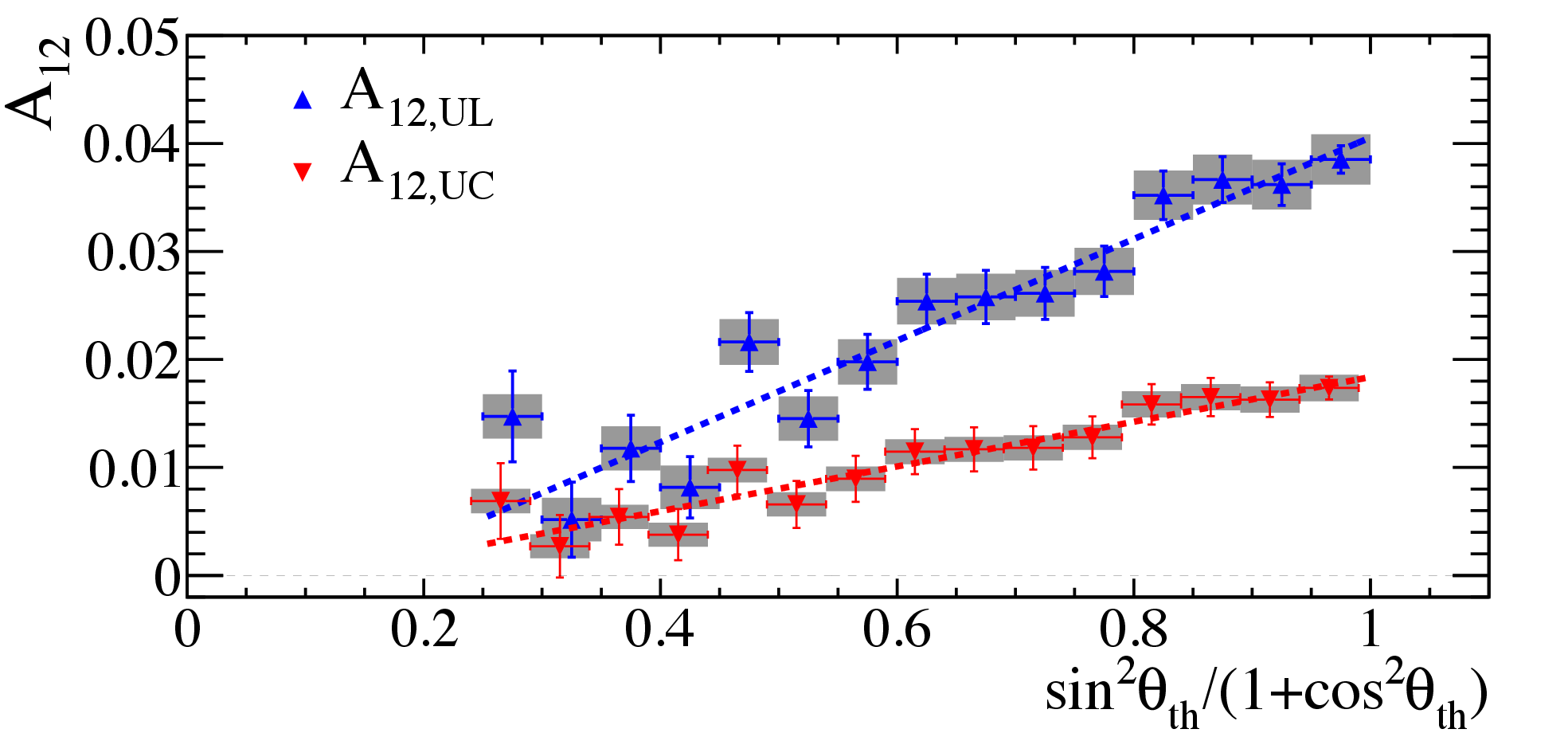}
 {\boldmath
      \put(-160,108){{(a) BaBar Preliminary}}
    }
  \includegraphics[width=0.48\textwidth] {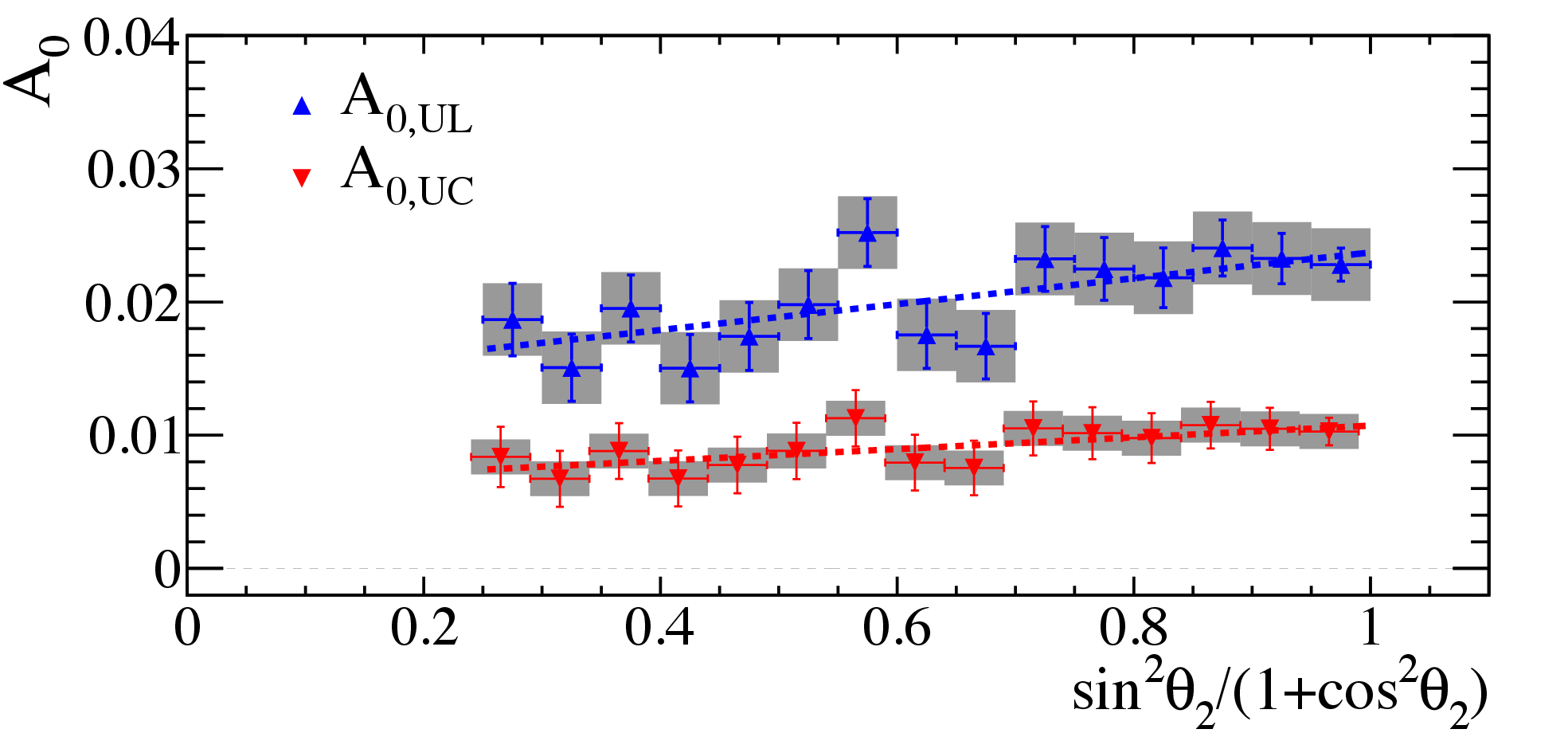}
   {\boldmath
      \put(-160,108){{(b) BaBar Preliminary}}
    }
\caption{ Collins asymmetries $vs.$ polar angle $\theta_{th}$ (a), and $\theta_2$ (b).
Blue and red triangles indicate the UL and UC double ratio, respectively,
while systematic uncertainties are shown by the gray bands.
The linear fit to $p_0+p_1\cdot x$ is represented by a dotted line
for each double ratio.
}
\label{fig:thetaRes}
\end{figure}

We study the behavior of the Collins asymmetries in the RF12 and RF0 frames
 as a function of pion
fractional energy $z$, pion transverse momentum $p_t$, and as
a function of the polar angle of the analysis axis.
Figure~\ref{fig:results} shows the corrected asymmetries in the RF12 frame
after subtraction of the background contributions,
  as an example. The systematic errors are represented by
error bands superimposed on the data points.
We observe a strong  increase of the asymmetry as a function
of pion fractional energy $z$ (Fig.~\ref{fig:results}(a)), which is in overall
good agreement with previous Belle results~\cite{belle}, and 
theoretical expectations~\cite{efremov}.
No previous data from $e^+e^-$ annihilation are available
for the asymmetries as a function of $p_t$.
This dependence was studied only in the space-like region 
at low $|Q^2|$ ($\sim 2.4$ (GeV/c)$^2$) in 
SIDIS experiments~\cite{sidis}, and so can be used to
investigate the evolution of the Collins function.
Finally, the Collins asymmetries are shown in Fig.~\ref{fig:thetaRes} as a function 
of the polar angle of the thrust axis $\theta_{th}$ in the RF12 frame, or the polar angle 
of the momentum of the second pion $\theta_2$ in the RF0 frame.
The dotted lines represent the results of the fit of a linear function, ($p_0+p_1\cdot x$),
to the data points.
In the case of RF12 the fitted lines extrapolate rather close to the origin,
which is consistent with the expectation.
In contrast, the fits clearly favor a non-zero constant parameter 
for the asymmetries in RF0;
this behavior may be explained by the fact that
$\theta_2$ is more weakly correlated
to the original $q\bar{q}$ direction than is the thrust axis.
These results are also in good agreement
with Belle data. \\

\vspace{-0.2cm}
In summary, we have reported preliminary BaBar results on the Collins asymmetries
in the pion system,  obtained from a data sample corresponding to an integrated luminosity 
of $468$ fb$^{-1}$ collected at a center-of-mass energy near
10.6 GeV.
The results shown are in good agreement with the Belle measurements,
and with theoretical expectations.  
In addition, we have extended the analysis by studying the asymmetry  
for different $p_t$ intervals.
This may help to shed light on the evolution of the Collins 
fragmentation function.
These data, combined with SIDIS and Belle data,
can also be valuable for improving global analyses,
such as that of Ref.~\cite{global}.

We thank Ralf Seidl of the Belle Collaboration for
useful discussions about the analysis method.


\begin{thebibliography}{99}
\bibitem{collins} J.~C.~Collins, \emph{Nucl. Phys. B} \textbf{396}, 161 (1993).
\bibitem{sidis}
A.~Airapetian (HERMES Collaboration),
 \emph{Phys. Rev. Lett.} \textbf{94}, 012002 (2005);\\
 E.~Ageev (COMPASS Collaboration),
 \emph{Nucl. Phys. B} \textbf{765}, 31 (2007).
\bibitem{boer}
D.~Boer, \emph{Nucl. Phys. B} \textbf{806}, 23 (2009). 
\bibitem{belle}
R.~Seidl (Belle Collaboration), \emph{Phys. Rev. Lett.} \textbf{96}, 232002 (2006);\\
R.~Seidl (Belle Collaboration), \emph{Phys. Rev. D} \textbf{78}, 032011 (2008).
\bibitem{global}
M.~Anselmino, M.~Boglione, U.~D'Alesio, A.~Kotzinian, F.~Murgia, A.~Prokudin, and C.~T\"urk,
 \emph{Phys. Rev. D} \textbf{75}, 054032 (2007).
\bibitem{babar}
B.~Aubert \emph{et al.} (BaBar Collaboration), 
\emph{Nucl. Instr. Meth. Phys. Res. A} \textbf{479}, 1  (2002).
\bibitem{thrust}
E.~Farhi, \emph{Phys. Rev. Lett.} \textbf{39}, 1587 (1977).
\bibitem{jetset}
T.~Tj\"{o}strand, \emph{Comput. Phys. Commun.} \textbf{82}, 74 (1994).
\bibitem{efremov}
A.~V.~Efremov, K.~Goeke, and P.~Schweitzer, \emph{Phys. Rev. D} \textbf{73}, 094025 (2006).

\end{thebibliography}
\end{document}